# Are diffusion models ready for materials discovery in unexplored chemical space?

Sanghyun Kim,[1†§] Gihyeon Jeon,[1†*] Seungwoo Hwang,[2] Jiho Lee,[2] Jisu Jung,[3] Seungwu Han,[2,4] and Sungwoo Kang[1,5*]

[1]Computational Science Research Center, Korea Institute of Science and Technology (KIST), Seoul 02792, Republic of Korea

[2]Department of Materials Science and Engineering, Seoul National University, Seoul, 08826, Republic of Korea.

[3]Climate&Energy R&D Group, Korea Institute of Ceramic Engineering and Technology (KICET), Jinju 52851, Republic of Korea.

[4]AI Center, Korea Institute of Advanced Study, Seoul, 02455, Republic of Korea.

[5]Division of Nanoscience and Technology, KIST School, University of Science and Technology (UST), Seoul 02792, Republic of Korea.

†These authors contributed equally.

§Present address: R&BD Center, Kumho Petrochemical, Yuseong-Gu, Daejeon 305-348, Republic of Korea.

*Correspondence: peulimieomy2@gmail.com (G.J.), sung.w.kang@kist.re.kr (S.Ka.)

While diffusion models are attracting increasing attention for the design of novel materials, their ability to generate low-energy structures in unexplored chemical spaces has not been systematically assessed. Here, we evaluate the performance of two diffusion models, MatterGen and DiffCSP, against three databases: a ternary oxide set (constructed by a genetic algorithm), a ternary nitride set (constructed by template informatics), and the GNoME database (constructed by a combination of both). We find that diffusion models generally perform stably in well-sampled chemical spaces (oxides and nitrides), but are less effective in uncommon ones (GNoME), which contains many compositions involving rare-earth elements and unconventional stoichiometry. Finally, we assess their size-extrapolation capability and observe a significant drop in performance when the number of atoms exceeds the trained range. This is attributed to the limitations imposed by periodic boundary conditions, which we refer to as the *curse of periodicity*. This study paves the way for future developments in materials design by highlighting both the strength and the limitations of diffusion models.

## INTRODUCTION

Advances in computational materials science have enabled rapid exploration of synthesizable materials via simulations such as density functional theory (DFT) calculations. The most straightforward approach involves screening candidate materials from existing databases, such as the Inorganic Crystal Structure Database (ICSD)[1] and the Materials Project.[2] However, such approaches are inherently limited to known compounds and cannot discover materials in uncharted chemical space. In fact, only 16% of ternary and 1% of quaternary compositions among the estimated possible combinations have been reported in the ICSD,[3] indicating that the existing materials library remains largely incomplete.

Crystal structure prediction (CSP) is a computational technique used to identify atomic configurations for a given composition, which is regarded as one of the grand challenges in materials science.[4] Most conventional CSP approaches fall into two main categories: direct search and template informatics (Scheme 1a,b). Direct search methods aim to explore the configuration space directly using algorithms such as random generation,[5] genetic algorithms,[6] and particle swarm optimization.[7] In these approaches, the local optimizations (relaxations) and energy evaluations of generated candidates are typically performed using DFT, which can be applied broadly without prior knowledge of the system. However, the high computational cost of DFT calculations limits the number of structures that can be explored, thereby restricting search of wide configuration space. In contrast, template informatics generate new structures by modifying known materials from existing databases, often by substituting elements with high-probability candidates suggested by machine-learning models.[8] Although DFT calculations are still required in this approach, they are only used to evaluate a small number of final candidate structures, significantly reducing the overall computational cost compared to direct search methods. Nevertheless, this strategy is fundamentally limited in its ability to discover entirely new structural prototypes, as it is inherently constrained by existing databases.[9] Recently, Merchant et al. proposed a complementary strategy that integrates these two approaches.[10] In their work, energy evaluations were performed using a deep learning-based energy prediction model, which enabled the screening of over 2.2 million hypothetical materials, known as GNoME database.

However, only 100 structures were generated per composition in their random structure searches, whereas typical CSP studies often require at least a thousand trials. This raises concerns about whether the newly discovered structures are truly representative of the ground state.

There has been growing interest in machine-learning interatomic potentials (MLIPs), which serve as surrogate models for DFT calculations to predict energies and forces, due to their higher computational efficiency and similar accuracy compared to DFT calculations.[11–13] MLIPs have also been employed to accelerate direct-search CSP.[14–16] However, a key challenge in applying MLIPs to CSP lies in the need to generate a training set before prediction, despite the lack of prior information about the target system. In our previous work, disordered structures generated from melt-quench DFT simulations were shown to be effective training data for MLIPs used in CSP.[17] Nevertheless, this method still requires considerable computational resources, taking approximately 3 to 4 days. In a follow-up study, we optimized the workflow to complete within 2 days by slightly compromising accuracy, and applied it to material discovery in ternary oxide systems.[18] However, this compromise in conditions resulted in notable prediction errors in a specific case, such as $RbRhO_2$. Therefore, it appears difficult to further reduce the prediction time using MLIPs without significantly affecting prediction accuracy. Moreover, these MLIP-based searches are generally applicable up to ternary compositions with fewer than 50 atoms per unit cell, and their reliability decreases for more complex multinary compositions or larger structures. This is due to the vastness of the configuration space of large multinary systems, which remains intractable for exhaustive exploration even with the acceleration provided by MLIPs in current structure search algorithms. These challenges underscore the need for more efficient CSP methods.

In recent years, generative models have been employed to discover novel materials with reasonable computational cost. A seminal study by Noh et al. introduced an image-based variational autoencoder (VAE) for representing materials and performing structural optimization in the latent space.[19] However, this method is typically trained on specific elemental systems and lacks generalizability. Moreover, it does not incorporate translational and rotational symmetries. To address these limitations, Xie et al. proposed the crystal diffusion variational autoencoder (CDVAE), which integrates a VAE with a diffusion model.[20] In their approach, atomic coordinates are directly diffused in real space and a denoising model with SE(3) equivariance is employed to naturally account for translational and rotational symmetries. Additionally, an element embedding scheme is adopted, allowing the model to handle a wide range of elements across the periodic table within a single model. Building upon similar approaches, several studies have further extended this framework to develop full diffusion models. Jiao et al. developed DiffCSP,[21] a model that performs diffusion over predefined atomic species and learns the corresponding reverse process to generate crystal structures. MatterGen, on the other hand, advances this concept by enabling diffusion not only over atomic positions and lattices but also over elemental types, thereby allowing end-to-end generation of materials including both chemical formula and structure.[22]

The development of universal MLIPs has further accelerated materials generations in conjunction with diffusion models. These models are trained on large-scale databases using a single architecture, enabling their application across a wide range of chemical systems.[23–26] Notably, they exhibit extrapolation capabilities across both compositional and structural spaces,[27,28] making them particularly valuable for materials discovery.[29] In fact, the previously mentioned work on MatterGen employed a universal MLIPs, MatterSim,[30] to perform structural relaxation and energy evaluation for the generated candidate structures. In addition, recent work has discovered novel ferroelectric materials by combining diffusion models with universal MLIPs.[31]

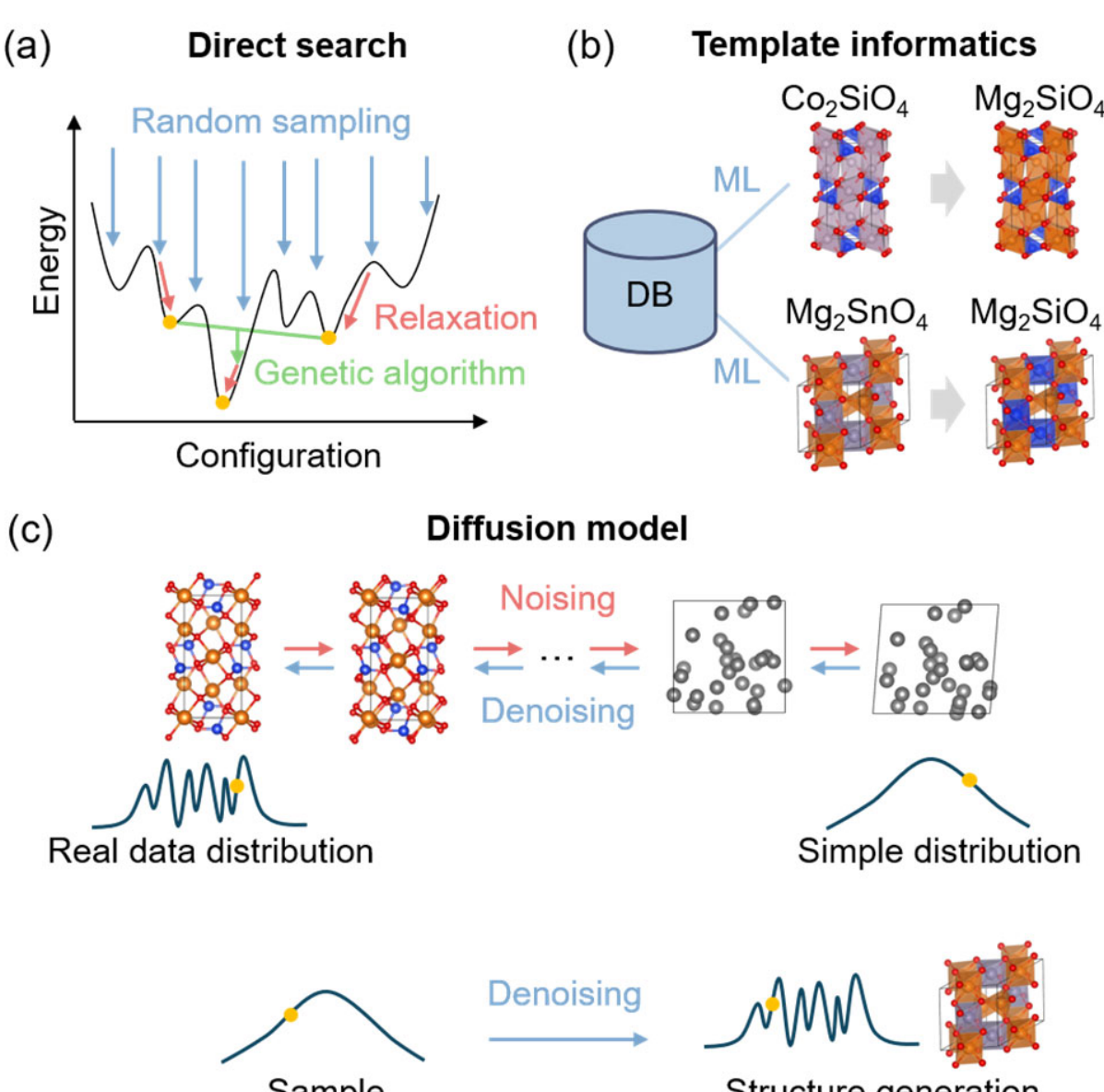

**Scheme 1.** Schematic illustrations of (a) direct searches, (b) template informatics, and (c) diffusion models.

The concept of diffusion models fundamentally differs from traditional CSP methods. Whereas conventional CSP aims to identify low-energy structures, diffusion models are designed to learn a generative process that maps a complex data distribution to a simple prior distribution (e.g., Gaussian). The goal is to generate novel structures that resemble the training data by sampling prior distribution (Scheme 1c). As such, there is no guarantee that structures generated by a diffusion model correspond to ground-state configurations. Instead, the generation relies on the assumption that, since the model is trained predominantly on low-energy materials, the generated samples from a similar distribution are also likely to have low energies. Indeed, ref. 32 reports a case where a diffusion model generates a structure that appears chemically reasonable but is in fact highly unstable when evaluated by its energy. To guide the generation toward structures with specific target properties, an adapter module can be incorporated into the diffusion model and fine-tuned, as implemented in MatterGen. Meanwhile, even with property guidance, the generated structures must exhibit low energies near the ground state to be considered synthesizable in inorganic materials design.[33] Yet, the performance of current diffusion models for CSP has not been systematically assessed. Note that most experimentally reported structures are already included in training sets of diffusion models. Consequently, rigorous testing requires databases of theoretically predicted structures with known ground-state (or near-ground-state) configurations. Constructing such reference sets typically involves CSP workflows and extensive DFT calculations, which are time-consuming and computationally demanding. As such, most existing benchmarks for diffusion models focus on evaluation metrics that are less time consuming, such as hull energy, novelty, and uniqueness.[21,22,34–36]

In this work, we systematically evaluate the performance of diffusion models, specifically DiffCSP and MatterGen, in identifying low-energy structures within a given chemical system, in a manner similar to the test suggested in ref. 32. To this end, we compare their performance using three established datasets, generated from well-known CSP methods, that consist of experimentally unexplored compositions: (1) the ternary oxide database generated using a genetic algorithm accelerated by MLIPs via the SPINNER code,[18] (2) the ternary nitride database constructed through template informatics,[37] and (3) the GNoME database.[10] Through this comparison, we assess the current capabilities and limitations of diffusion models in crystal structure prediction. In addition, we report a key technical challenge: the limited extrapolative ability of these models when applied to systems with a larger number of atoms not represented in the training data. We provide a theoretical explanation for the origin of this limitation.

## RESULTS

### Structure prediction with diffusion models and evaluation metrics

Here, we compare the performance of pretrained MatterGen and DiffCSP models. For MatterGen, we use a pretrained model trained on the Alex-MP-20 database, which combines structures with up to 20 atoms collected from the Alexandria[38] and Materials Project[2] databases. We employ a fine-tuned model capable of generating atomic structures within a specified chemical system (e.g., Sr–Ti–O). However, it does not directly specify the exact chemical formula (e.g., $SrTiO_3$). In this approach, the generation process begins from random atomic structures with unspecified atom types, and the element identities are progressively denoised so that the final composition emerges in the last step. For DiffCSP, we use a pretrained model trained on the MPTS-52 database,[39] which contains selected crystal structures from the Materials Project with up to 52 atoms. In contrast to the MatterGen, DiffCSP starts from random structures of given atom types, meaning that the chemical formula is explicitly provided as input.

To evaluate the performance of CSP methods and diffusion models, we calculate the hull energies of the candidate materials. Hull energy (or energy above the convex hull) quantifies the thermodynamic stability of a material with respect to decomposition into other compounds within a chosen reference set. Scheme 2 illustrates two types of hull energies. In the first case, $E_{MP}$, the Materials Project (MP) database is used as the reference. Because the MP database does not include the discovered candidate structures, some candidates may appear to have negative hull energies relative to this reference, implying that they would further stabilize the MP convex hull if they are incorporated. In the second case, $E_{All}$, the reference set is extended to include all structures discovered in this study, so the minimum energy is always zero by construction. $E_{MP}$ is used to evaluate whether a chemical system is stable relative to the existing database. $E_{All}$ is used to compare the relative stability among generated candidates within the same or different chemical formulas. As template informatics and MatterGen do not explicitly specify the composition and can generate diverse chemical formulas within a given chemical system, $E_{All}$ provides a fair basis for performance comparison with the genetic algorithm, DiffCSP, or between the two methods.

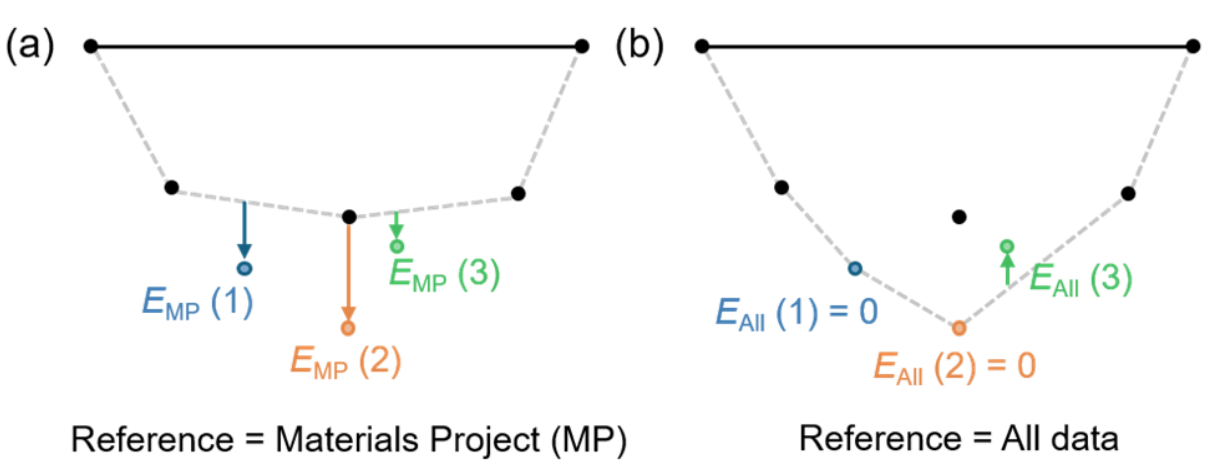

**Scheme 2.** Schematic illustration of hull energies defined by (a) the MP reference and (b) the all-data reference.

We generate 1024 structures for each chemical system with MatterGen. With DiffCSP, 1024 structures are generated for each formula, considering formula units from 1 to 4 (total 4048 structures). If the basic formula unit contains too many atoms such that the total number of atoms in the unit cell would exceed 60, we reduce the maximum formula unit accordingly. The generated structures are then optimized and evaluated using the universal MLIP, SevenNet-MF-ompa,[40] which achieved one of the highest scores in Matbench Discovery.[29] For materials within a certain energy window, we further optimize the structures and evaluate their energies with DFT calculations using the AMP[2] automation code.[41] Details of these procedures are provided in the Methods section.

### Comparison with genetic algorithm accelerated by MLIPs: Ternary metal oxide database

We first evaluate the performance of MatterGen and DiffCSP using the theoretical ternary oxide database[18] constructed by the SPINNER,[17,42] which integrates a genetic algorithm with MLIPs. In the referenced study, 181 oxide chemical systems were explored, leading to the discovery of 48 stable chemical formulas across 45 chemical systems. Note that in the SPINNER database, the most stable structures are taken from the R2SCAN functional[43] and compared with PBE results,[44] so an error of about ~30 meV/atom may exist.

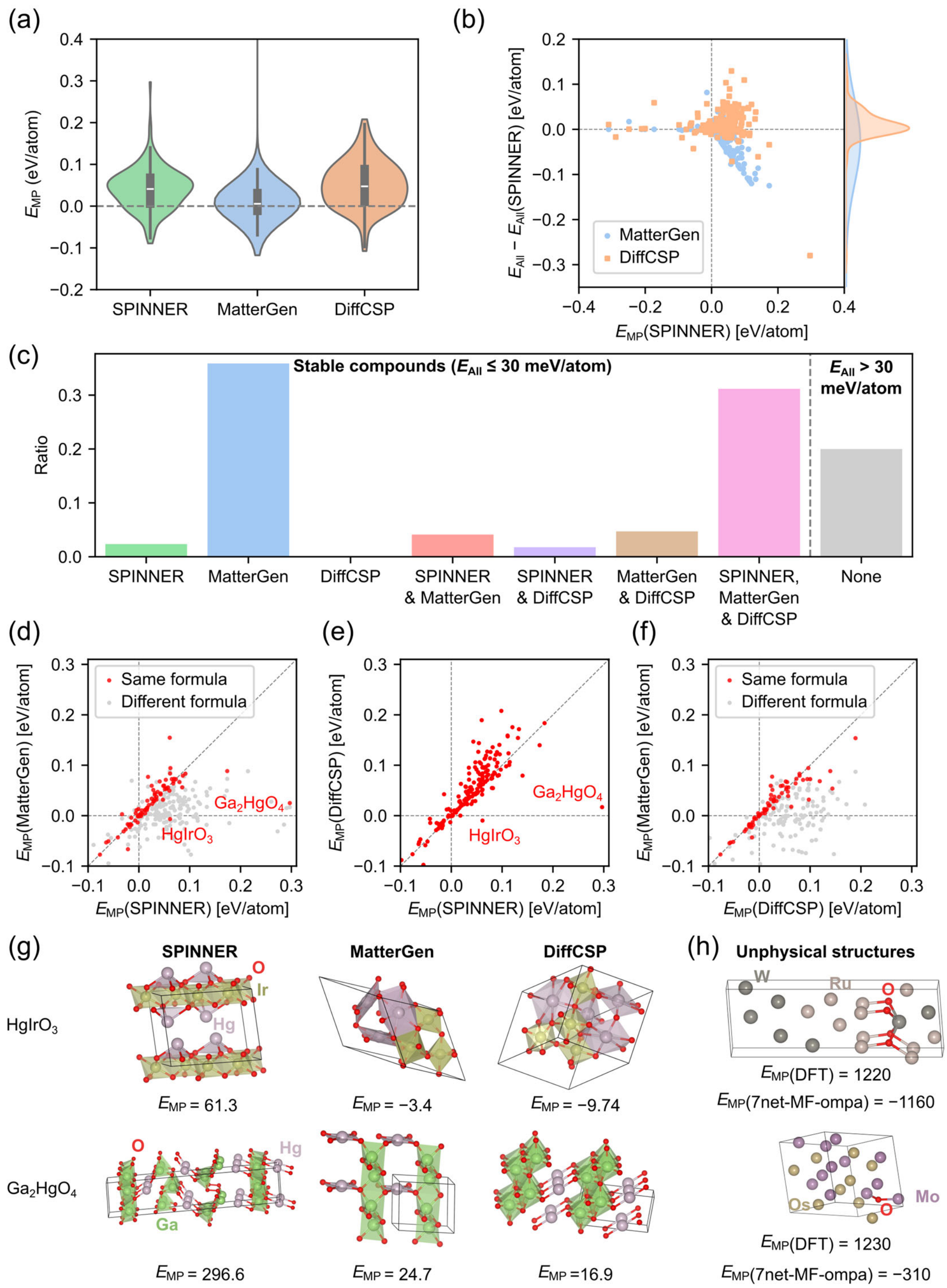

**Figure 1. Tests on the ternary metal oxide database generated by the genetic algorithm (SPINNER).** (a) Distribution of $E_{MP}$ for each method. (b) $E_{All}$ differences of MatterGen and DiffCSP relative to the reference structures, plotted against the $E_{MP}$ of the reference structures. (c) Fraction for identifying stable chemical systems by each method. If a single method is listed, it indicates that only that method discovers a stable structure in the given chemical system. If two or three methods are listed, the stable structures are identified by all of those methods. "None" indicates that no method finds a stable structure. The sum of these values is equal to 1. Comparison of $E_{MP}$ for the same and different chemical formulas between (d) MatterGen and SPINNER, (e) DiffCSP and SPINNER, and (f) MatterGen and DiffCSP. (g) Representative structures and their energies (meV/atom). (h) Representative unphysical structures generated by MatterGen.

Figure 1a shows the distribution of $E_{MP}$ for structures generated by each method. We find that MatterGen produces the largest fraction of structures with negative hull energies, followed by SPINNER and then DiffCSP. However, because $E_{MP}$ is referenced to the Materials Project, it does not account for compositional differences between the generated candidates. To address this, we compare $E_{All}$ in Figure 1b. Even in this comparison, MatterGen is also found to generate structures with lower energies than those from SPINNER. A more detailed analysis is shown in Figure 1c, highlighting which method identifies the lowest-energy structure in each chemical system. (30 meV/atom window is applied for considering the errors in the DFT methods.) MatterGen most frequently discovers stable compounds, followed by SPINNER and DiffCSP. This indicates that MatterGen is particularly effective at rapidly identifying stable oxide chemical systems by exploring diverse chemical formulas, whereas the other two methods are limited to predefined ones.

We next investigate whether the stable compounds discovered by MatterGen correspond to the same chemical formulas as those identified by SPINNER, or to different formulas. To this end, we compare the $E_{MP}$ values of MatterGen and SPINNER for both identical and distinct compositions in Figure 1d. For the same formulas, the results from MatterGen align relatively well with those of SPINNER, indicating that both methods generally produce structures with similar hull energies. However, for different compositions, MatterGen frequently predicts more stable compounds, suggesting that most of the apparent advantage of MatterGen in Figures 1a-c arises from generating diverse compositions. Surprisingly, even for the same composition, MatterGen occasionally predicts much lower-energy structures than SPINNER. For example, in $IrHgO_3$, SPINNER yields a structure with 61.3 meV/atom, whereas MatterGen finds a −3.4 meV/atom structure. Similarly, for $Ga_2HgO_4$, SPINNER produces a 296.6 meV/atom structure, while MatterGen generates a 24.7 meV/atom structure, which falls within the range considered synthesizable. When comparing DiffCSP with SPINNER (Figure 1e) and with MatterGen (Figure 1f), we observe a similar trend: all methods exhibit comparable performance when restricted to the same chemical formula.

To investigate the origin of SPINNER's failure in $HgIrO_3$ and $Ga_2HgO_4$, we compare the atomic structures of generated by SPINNER, MatterGen, and DiffCSP (Figure 1g). For $HgIrO_3$, SPINNER predicts a structure in which Hg forms 3-fold polyhedra with O atoms, whereas MatterGen and DiffCSP both generate structures with 6-fold polyhedra. In these latter cases, the O atoms within the polyhedra are arranged in nearly planar configurations. In $Ga_2HgO_4$, SPINNER yields linear Hg–O dumbbell units, while MatterGen produces four-coordinated planar Hg environments. In addition, Ga atoms form five-fold pyramidal units in the SPINNER structure but appear as six-fold distorted polyhedra in the DiffCSP structure. Since these target formulas contain only 20 and 7 atoms ($HgIrO_3$ and $Ga_2HgO_4$, respectively), a size regime that the genetic algorithm can generally handle well, the failure of SPINNER arises not from limitations of the structure generation process but from the inability of the MLIPs to accurately describe the energies of such local motifs. Specifically, the MLIPs employed in SPINNER were trained using a melt–quench–annealing dataset, but both the system sizes and simulation times used in ref. 18 were smaller than those in the original protocol described in ref. 17. In addition, local environments such as 6-fold and 4-fold planar Hg–O polyhedra are highly anisotropic and thus particularly difficult to sample within short MD trajectories, further limiting their representation in the training data.

We report a few cases where candidate structures suggested by MLIPs are found to be unstable when evaluated with DFT. Representative examples are shown in Figure 1h. In these cases, the $E_{MP}$ values calculated with DFT exceed 1 eV, whereas those obtained with SevenNet-MF-ompa are highly negative. This discrepancy likely arises because the structures correspond to unconventional compositions with too few oxygen atoms. As such compositions are absent from the MLIP training set, the predicted energies become unphysical. (Note that the results shown in Figures 1a–f have been filtered to exclude these unphysical cases.) We note that such artifacts can be screened out by considering common oxidation numbers, which provides a practical guideline for applying diffusion models in materials design.

**Comparison with template informatics: Ternary nitride database**

We evaluate the performance of MatterGen and DiffCSP on the ternary nitride database constructed by the template informatics.[37] We randomly select 63 chemical systems with $E_{MP} = 0$ and 78 systems with $E_{MP} > 0$ (a total of 141), excluding those included in the MP-20 database. MatterGen is applied across all of these systems and DiffCSP is applied to predict structures at the most stable (lowest $E_{MP}$) chemical formula at each selected system. Figure 2a shows the $E_{MP}$ distributions predicted by these methods. Similar to the test on the ternary oxide database, MatterGen generates the largest fraction of negative-hull-energy materials. In Figure 2b, we compare $E_{All}$ values and find that both MatterGen and DiffCSP produce more stable structures than those obtained from template informatics. Figure 2c summarizes which method identifies the lowest-energy structure in each chemical system. We observe that the two most frequent cases are: (i) MatterGen alone discovers a stable chemical system, and (ii) all methods identify the stable system. These results indicate that MatterGen is the most effective approach for finding the most stable chemical systems.

Figures 2d–f compare the $E_{MP}$ values obtained from the different methods for both identical and distinct chemical formulas. In comparison with template informatics (Figure 2d), MatterGen generally predicts lower energies for the identical chemical formulas. Similarly, DiffCSP identifies more stable structures than template informatics (Figure 2e). The direct comparison between DiffCSP and MatterGen shows that they predict structures with nearly identical energies for the same chemical formulas (Figure 2f).

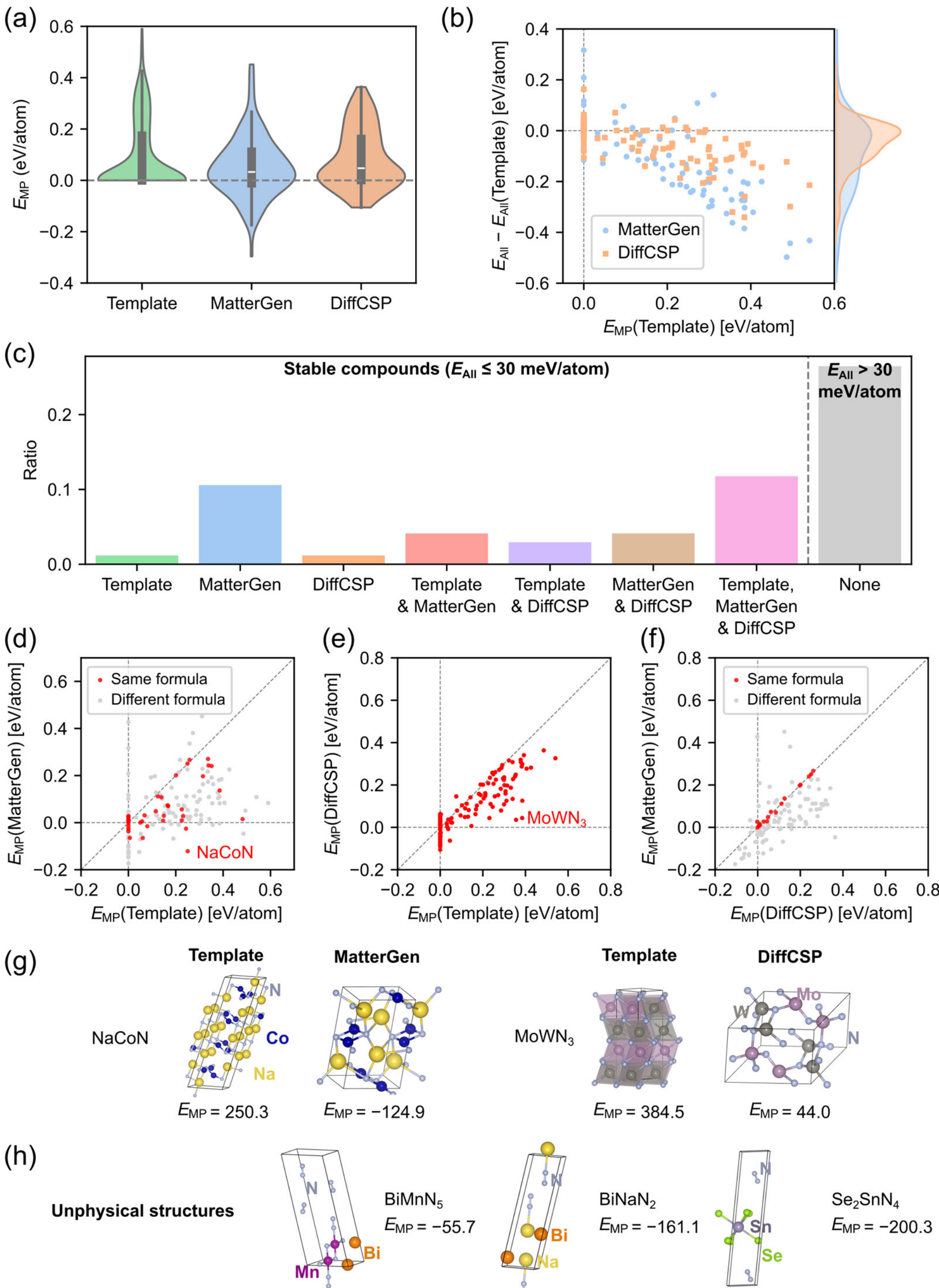

**Figure 2. Tests on the ternary nitride database generated by the template informatics.** (a) Distribution of $E_{MP}$ for each method. (b) $E_{All}$ differences of MatterGen and DiffCSP relative to the reference structures, plotted against the $E_{MP}$ of the reference structures. (c) Fraction of cases identifying stable chemical systems. If a single method is listed, it indicates that only that method discovers a stable structure in the given chemical system. If two or three methods are listed, the stable structures are identified by all of those methods. "None" indicates that no method finds a stable structure. The sum of these values is equal to 1. Comparison of $E_{MP}$ for the same and different chemical formulas between (d) MatterGen and SPINNER, (e) DiffCSP and SPINNER, and (f) MatterGen and DiffCSP. (g) Representative structures and their energies (meV/atom). (h) Representative unphysical structures generated by MatterGen.

In some cases, template informatics generates significantly unstable structures. For example, in NaCoN, template informatics yields a structure with 250.3 meV/atom, whereas MatterGen generates a structure with –124.9 meV/atom (Figure 2g). Likewise, for $MoWN_3$, template informatics predicts 384.5 meV/atom, while DiffCSP produces 44.0 meV/atom. To further examine these cases, we analyze the structural prototypes using AFLOW XtalFinder.[45] We find that the lowest-energy structure of NaCoN does not match any existing prototype identified by XtalFinder. This suggests that such compositions could not be predicted by template informatics in the first place. In contrast, diffusion models are able to predict stable structures even when the exact prototypes are absent from the training dataset.

We also report that unphysical structures are generated in numerous systems, as shown in Figure 2h. Unlike in the oxide case, the unphysical structures in nitrides often include the formation of $N_2$ molecules. Interestingly, these structures exhibit low DFT energies, which contrasts with the oxide cases. This occurs because $N_2$ gas is highly stable, making the overall structure appear stable. (In the main results of Figures 2a-f, we manually excluded such cases by inspecting the N–N bond distances.) We further find that many of these artifacts originate from unphysical oxidation states generated by MatterGen, as in Figure 2h. Thus, applying constraints based on reasonable oxidation numbers can filter out a significant fraction of these cases. However, even when using DiffCSP, where the chemical formula is fixed and common oxidation states are applied, structures containing $N_2$ molecules are still occasionally generated. These results highlight that empirical filtering remains necessary to address such artifacts, and further improvements to the generation models are needed.

**Comparison with the GNoME database**

For the final comparison test, we generate structures for the same chemical systems and chemical formulas using MatterGen and DiffCSP, respectively, for materials listed in the GNoME database.[10] A total of 100 chemical formulas are selected. Among these, 50 formulas are chosen deliberately: 20 ternary, 20 quaternary, and 10 quinary systems, with a focus on chemical formulas containing anions and semimetals, since purely metallic systems generally form simple crystal structures. The remaining 50 formulas are randomly selected.

For the first set of 50 formulas, we further select 17 ternary and quaternary materials at random and perform a stress test by comparing SPINNER and GNoME, as shown in Figure 3a. In most cases, SPINNER produces structures with similar energies to those in GNoME. However, it also identifies significantly more stable structures for $Yb_4Ta_2O_9$ (222.5 meV/atom lower) and $CeNpSbN_2$ (94.8 meV/atom lower).

We then test DiffCSP and MatterGen for the 100 chemical systems, as shown in Figures 3b,c, respectively. To reduce the computational cost, we develop a systematic protocol in which most calculations are performed using MLIPs, and only low-energy candidates are subsequently evaluated with DFT (see Methods section for details). For DiffCSP, the ternary and quaternary compositions yield energies similar to those from GNoME, whereas for quinary materials, DiffCSP tends to produce higher-energy structures. For $Yb_4Ta_2O_9$ and $CeNpSbN_2$, DiffCSP generates much more stable structures compared to GNoME, consistent with the results from SPINNER. However, when compared with SPINNER, the energy of the $Yb_4Ta_2O_9$ structure from DiffCSP is 100.2 meV/atom higher. Moreover, DiffCSP predicts an $E_{\mathrm{MP}}$ of 260.8 meV/atom for $KTa_5O_8$, while GNoME yields –1.2 meV/atom (see Figure 3f for the atomic structures). Based on these results, DiffCSP performs worse on the GNoME database than on the oxide and nitride databases described above. We attribute this poor performance partly to the inclusion of rare-earth elements, such as Yb, which are underrepresented in existing experimental databases. However, DiffCSP also fails for $KTa_5O_8$, even though its elements are not uncommon. We attribute this to the unusual cation ratio of 1:5, which is rarely represented in experimental databases, as the cation ratio similar to 1:1 is abundant in the ICSD database.[18]

For MatterGen, most of the structures yield energies similar to those from GNoME (Figure 3c). In particular, for quinary compositions, all structures generated by MatterGen have $E_{\mathrm{All}}$ = 0 eV. To further evaluate the performance of MatterGen, we separate the comparison into cases with the same chemical formula and with different formulas, as shown in Figure 3d. As can be seen, in the quinary systems, there are no cases where MatterGen generates structures with the same chemical formula as GNoME. In other words, all generated structures differ from those of GNoME, yet still lie on the convex hull. This may be because, for compositions involving rare-earth elements, the binary, ternary, and quaternary phase diagrams are not yet fully established, and therefore neither the GNoME nor the MatterGen structures can be confidently regarded as the true ground state. Further rigorous investigation will be necessary in future studies to clarify this point.

In the algorithm of MatterGen, there is no strict constraint on the chemical system; rather, the generated structures result from fine-tuning a pretrained model against existing databases. As a result, some generated structures may not actually belong to the specified chemical system. To quantify this, we investigate the ratio of generated structures that satisfy the given chemical system, as shown in Figure 3e. For example, if 512 structures out of the 1024 generated for a chemical system satisfy the specified chemical system, then the success ratio is 0.5. We find that the success ratio decreases from ternary to quinary systems, with all quinary chemical systems showing values below 0.5. Moreover, many chemical systems exhibit a success ratio of 0, as indicated in the lower part of Figure 3e. These cases predominantly involve radioactive or rare-earth elements, further supporting our argument that diffusion models are undertrained in such rare chemical spaces.

Figure 3f shows representative atomic structures where GNoME or the diffusion models fail to predict stable materials. For $Yb_4Ta_2O_9$ and $CeNpSbN_2$, GNoME fails to identify the stable structures. The $Yb_4Ta_2O_9$ structure is not classified into any known prototype and is therefore likely generated from random structure search. In GNoME, only about 100 random samples are generated for each composition, which is often insufficient to guarantee the true ground state, leading to unsuccessful prediction in this case. $CeNpSbN_2$ is known to share the same prototype as $Th_2SbN_2$ (ICSD ID = 16062), with Th partially substituted by Ce and Np. However, it is unclear whether this structure in GNoME originates from template informatics or random structure search. In the Pr–Ir–Ge–I system, the local motifs of the structures predicted by GNoME and MatterGen are similar, and DiffCSP yields energies comparable to those from GNoME. Thus, GNoME does not seem to fail for this chemical formula.

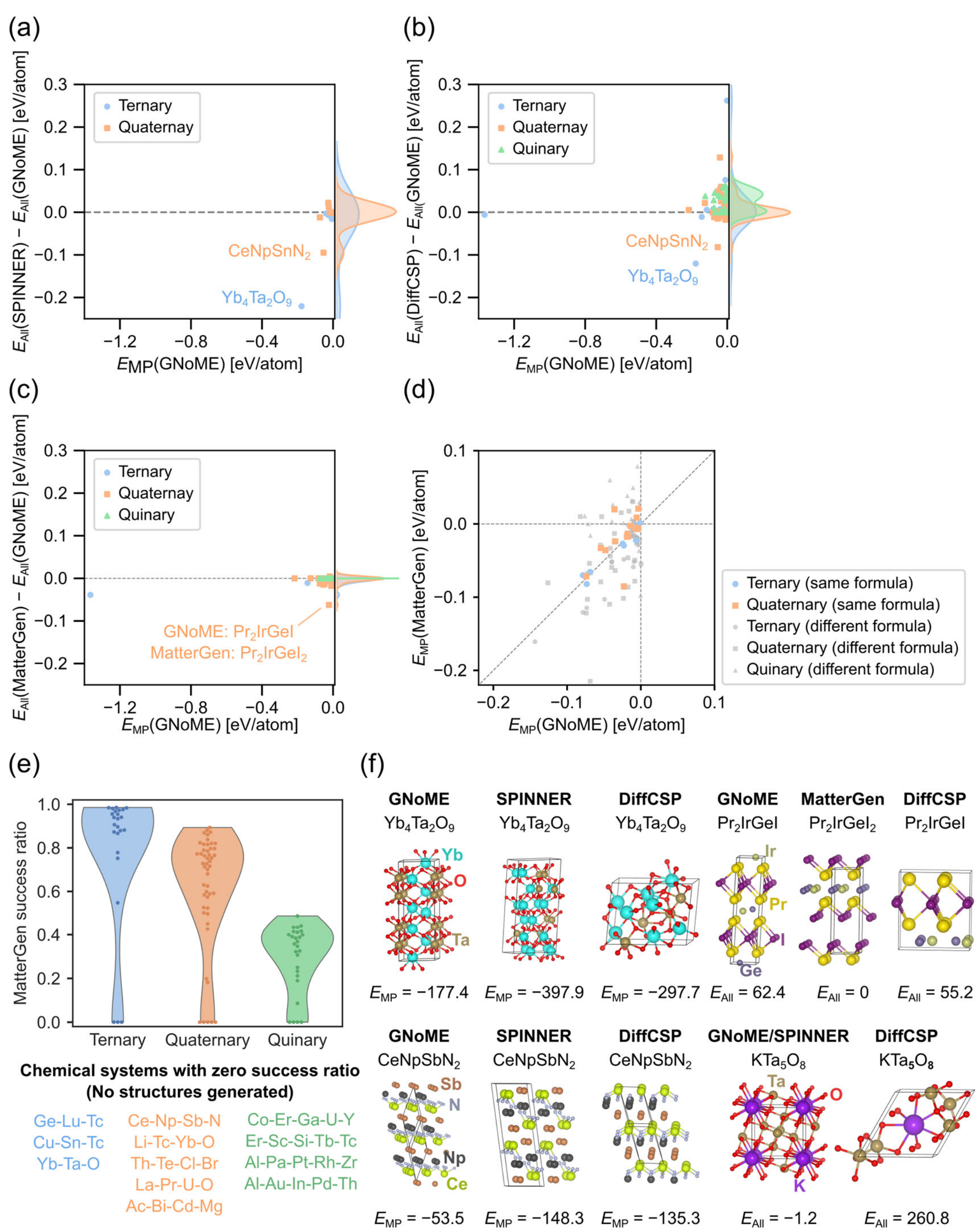

**Figure 3. Test on the GNoME database.** Relative $E_{All}$ of generated structures compared to the GNoME reference structures for (a) SPINNER, (b) DiffCSP, and (c) MatterGen. (d) $E_{MP}$ of GNoME structures compared to the MatterGen-generated structures in the same and different chemical formulas. (e) Upper part: The success ratio of MatterGen. Lower part: Chemical systems with zero success rate. (f) Representative structures and their energies (meV/atom).

Instead, MatterGen discovers a more stable structure in a different chemical formula within the same chemical system.

To investigate whether the reduced performance on the GNoME database is primarily due to compositional under-sampling or the prevalence of structurally novel configurations, we analyzed the test structures using AFLOW-XtalFinder[45] to determine their correspondence to known prototypes. To account for partially substituted structures (also see ref. 46), we performed prototype matching by grouping cations or anions into identical species and testing all possible combinations of these grouped elements for lower-order prototype equivalence. In addition, all structures were symmetrized using phonopy[47,48] prior to prototype matching to ensure consistency in symmetry detection. Quinary compositions are excluded because their symmetries are often severely distorted, which can prevent reliable prototype identification even when an underlying prototype may exist. We find that the only 18 out of 72 (20%) ternary and quaternary materials do not correspond to any known prototype (see Table S1). In addition, we suspect that, even in this 20% cases, many structures seem to be originated from known prototypes, but are not identified by AFLOW-XtalFinder as such due to symmetry breaking induced by partial substitutions. (However, we do not have quantitative evidence to confirm this, and more rigorous methodologies for prototype analysis will be required in future studies.) Accordingly, the reduced performance of diffusion models on GNoME is more plausibly attributed to compositional under-sampling than to a lack of structural coverage in the training data. (For comparison, for the ternary oxide database, 24 out of 48 stable oxide structures correspond to previously unreported prototypes[18] and the nitride dataset was constructed through ionic substitution,[37] and thus most of the structures can be considered to have existing prototypes.)

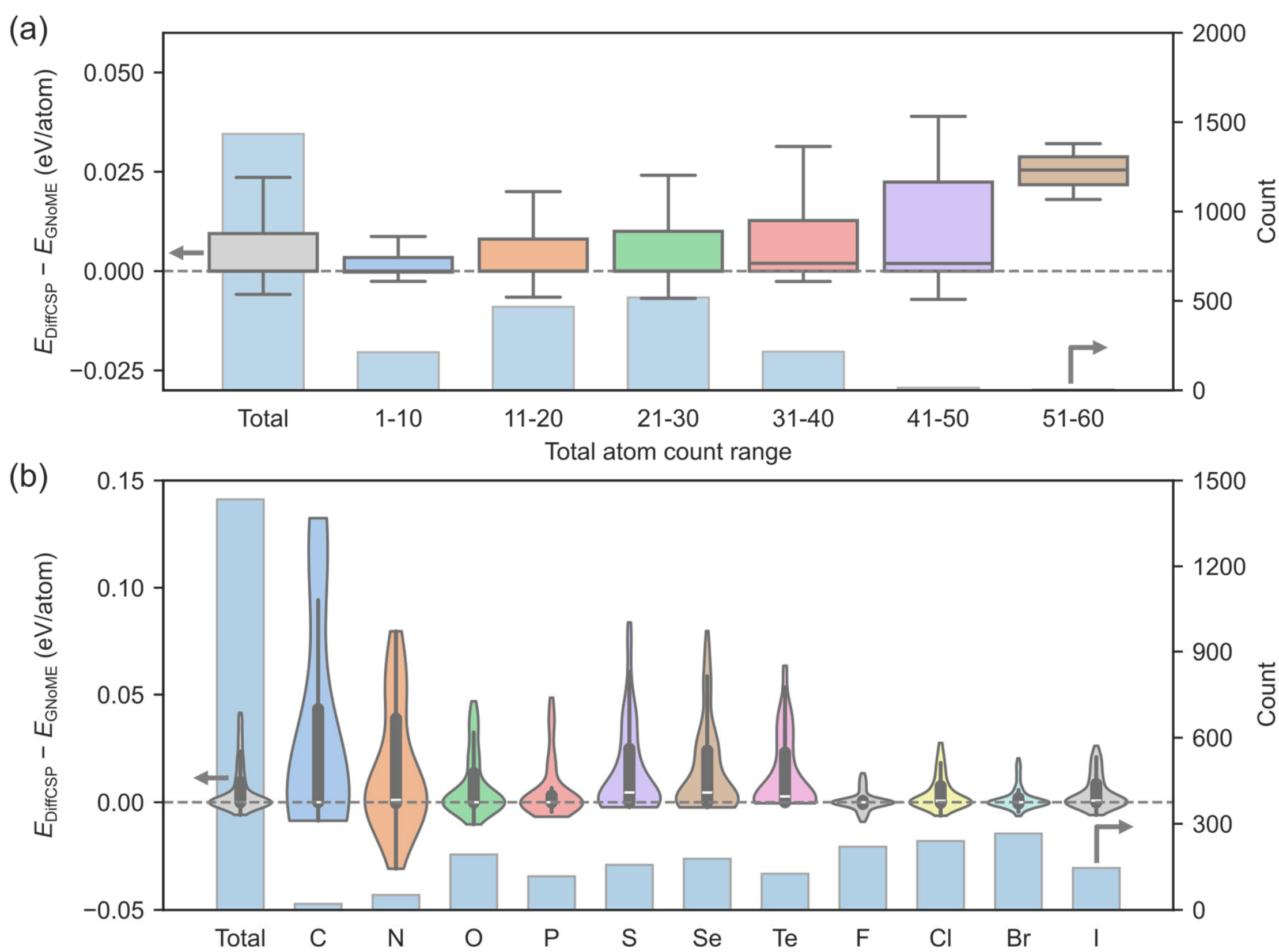

**Figure 4. Large-scale analysis of ternary compounds in GNoME database.** Distribution of the energy differences between the minimum-energy structures generated by DiffCSP and the corresponding GNoME structures, shown as a function of (a) the number of atoms in the unit cell as a box plot and (b) the included element types as a box plot.

From the above results, we identify that diffusion models struggle to accurately predict rare-earth–containing and quaternary materials. Accordingly, we carry out further large-scale evaluations on ternary systems that do not exhibit these features to assess diffusion-model performance in a broader dataset. Specifically, we select 1,434 ternary compositions from the GNoME database that (i) do not contain rare-earth elements, (ii) include at least one anion, and (iii) exhibit chemically reasonable oxidation states. Due to the large number of compositions, DFT calculations are not performed for this test. Instead, energies are evaluated using SevenNet-MF-ompa only, and a single formula unit identical to the reference structure is considered for each composition. Accordingly, this analysis is intended to reveal systematic trends across material classes and compositions rather than to identify precise ground states or individual outliers.

Figure 4a shows box plots of the energy differences between structures generated by DiffCSP and the corresponding reference structures from the GNoME database as a function of the number of atoms in the unit cell. The number of occurrences for each element is also presented as bar charts. As noted above, since DFT calculations are not performed for this large-scale test, the results may contain outliers arising from the intrinsic errors of SevenNet-MF-ompa. To mitigate this effect, we exclude the top and bottom 10% of data points from the plots. The resulting distributions show that, in most cases, DiffCSP generates structures within 50 meV/atom of the GNoME reference structures, indicating a reasonable level of accuracy similar to the previous tests. In addition, it shows that the decreasing performance when increasing number of atoms.

Figure 4b presents the element-resolved performance of DiffCSP in terms of energy prediction. With the exception of carbides, most generated structures fall within 0.05 eV/atom of the reference energies. In contrast, carbide systems exhibit a larger fraction of higher-energy structures. We speculate that this behavior arises from the chemical complexity of carbides, which can exhibit a wide range of oxidation states and may involve both anionic and cationic character, combined with their relatively limited representation in the training dataset. Notably, for nitrides, DiffCSP frequently identifies structures with lower energies than those reported in the GNoME database. This observation suggests that the existing prototype libraries for nitride systems may be incomplete (as GNoME structures are speculated to be largely generated by ion substitutions as discussed above; and see ref. 46) and that diffusion-based approaches may help uncover additional low-energy configurations in this chemical space.

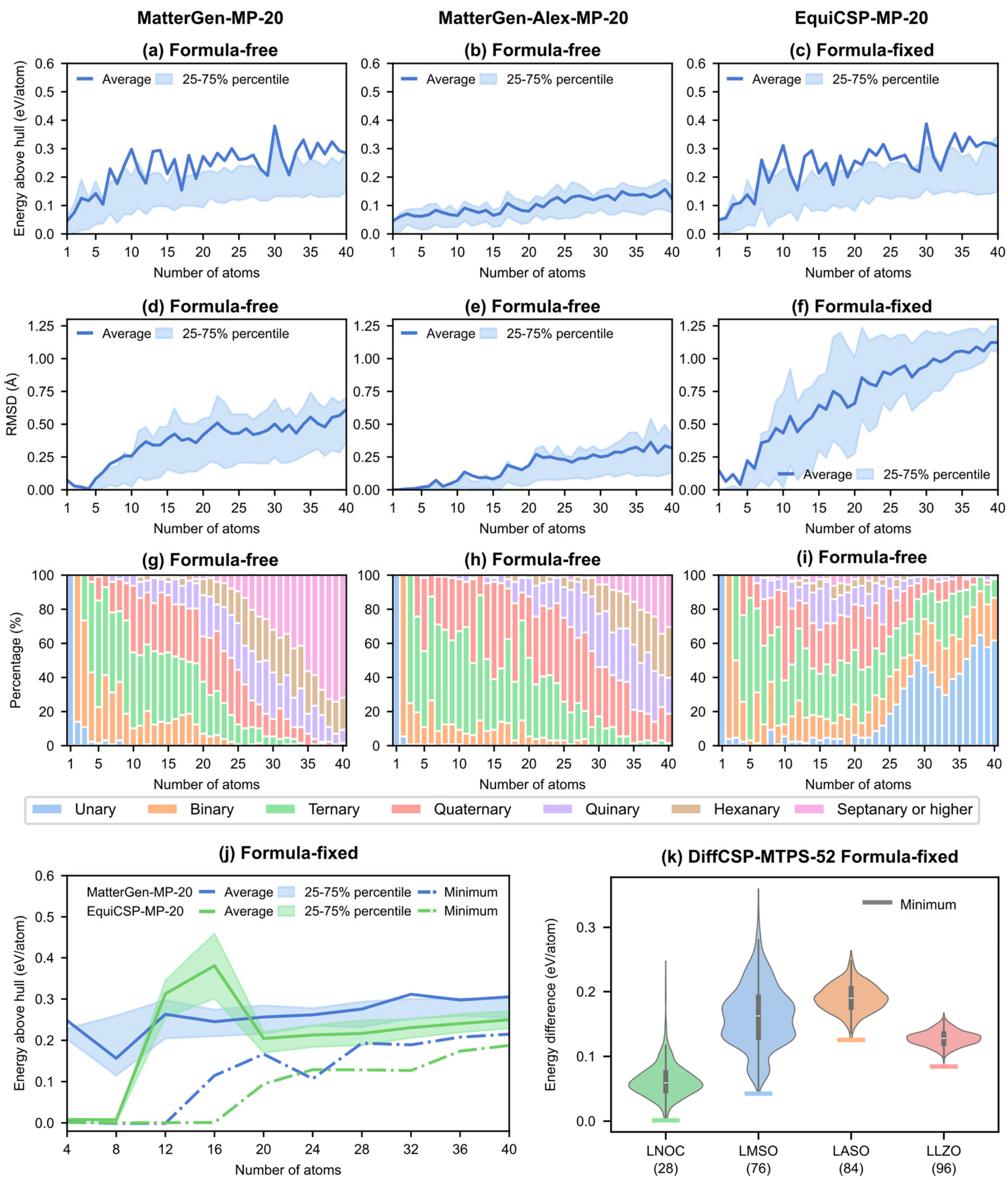

**Figure 5. Size-extrapolation capability of diffusion models.** (a-c) $E_{MP}$ of generated structures: results for MatterGen trained on (a) MP-20 and (b) Alex-MP-20 with chemical-formula-free settings, and (c) EquiCSP trained on MP-20 with chemical-formula-fixed settings. (d-f) Root-mean-square displacement (RMSD) of generated structures: results for MatterGen trained on (d) MP-20 and (e) Alex-MP-20 with chemical-formula-free settings, and (f) EquiCSP trained on MP-20 with chemical-formula-fixed settings. (g-i) Composition type ratios of structures generated by (g) MP-20-trained MatterGen, (h) Alex-MP-20-trained MatterGen, and (i) MP-20-trained EquiCSP. (j) $E_{MP}$ of $CuAlO_2$ at different formula units (i.e., different numbers of atoms per unit cell) generated by MP-20-trained MatterGen and EquiCSP. (k) Performance of DiffCSP on solid-state electrolyte materials. Numbers in parentheses indicate the number of atoms in the unit cell. LNOC: $LiNbOCl_4$, LMSO: $Li_2Mg_2S_3O_{12}$, LASO: $LiAlSiO_4$, and LLZO: $Li_7La_3Zr_2O_{12}$.

### Size extrapolation capability

We now investigate the size-extrapolation capability of diffusion models. We use the pretrained MatterGen model trained on structures with fewer than 20 atoms, and the pretrained

DiffCSP model trained on structures with fewer than 52 atoms. In addition to these models, we evaluate the EquiCSP,[35] which is designed to guarantee lattice permutation, in contrast to DiffCSP. For a fair comparison with MatterGen models, we use the pretrained EquiCSP model trained on the MP-20 database. We monitor the performance of these models when they generate structures with atom counts beyond their respective training ranges.

Figures 5a-c show the hull energies of structures generated by MatterGen trained with MP-20 (MatterGen-MP-20, henceforth), MatterGen-Alex-MP-20 and EquiCSP-MP-20, respectively, as a function of atom number ($N_{atom}$). For MatterGen, while the original code randomly selects $N_{atom}$ based on the training-set distribution, we modify the code to define $N_{atom}$ explicitly in the unit cell and generate 100 structures for each $N_{atom}$ value. For EquiCSP, the formula-fixed model (stated as CSP-mode in the original paper[35]) is conditioned to generate structures via the same compositions as the structures generated from MatterGen-MP-20. For all models, the energy above hull increases significantly as $N_{atom}$ grows. This indicates that these generative models extrapolate poorly in the untrained $N_{atom}$ region.

On the other hand, it remains unclear whether the generation of higher-energy structures at large $N_{atom}$ is due to a fundamental degradation in the performance of diffusion models or arises from the extreme increase in the complexity of the configuration space—which leads to a higher likelihood of producing metastable structures. Therefore, we analyze the root-mean-square displacement (RMSD) of the generated structures by comparing before and after optimization (Figures 5d-f). RMSD also increases with $N_{atom}$, indicating that the generated structures deviate more strongly from local minima and that model quality degrades with increasing size. This suggests that the structures generated by MatterGen and EquiCSP deviate from equilibrium positions, indicating that the model's overall structural fidelity decreases with size. We also note that EquiCSP shows worse extrapolation performance compared to MatterGen models.

Figures 5g-i show the number of elements for the MatterGen-MP-20, MatterGen-Alex-MP-20, and EquiCSP-MP-20, respectively, with Figure S1 providing the training-set reference. Here, we use formula-unfixed version of EquiCSP (stated as *ab initio* structure generation the original paper[35]). We find that the number of elements tends to increase with $N_{atom}$. Notably, structures with six or more elements are almost absent in the original training dataset (Figure S1), yet their proportion rises significantly once $N_{atom}$ exceeds 20 in both MatterGen models (Figures 5g, h). In contrast, for EquiCSP (Figure 5i), the proportion of unary structures increases sharply in the extrapolation regime ($N_{atom} > 20$). This indicates a reduced capability of EquiCSP to generate chemically diverse or novel structures in the untrained domain, suggesting that enforcing lattice permutation and periodic translation equivariance is insufficient for ensuring extrapolation capability. Therefore, these results collectively support the conclusion that the quality of diffusion models deteriorates at large $N_{atom}$ compared to small $N_{atom}$.

**Scheme 3.** Schematic illustration comparing the architectures of (a) Machine-learning interatomic potentials (MLIPs) and (b) diffusion models. (a) MLIPs derive the total energy $E_{tot}$ by summing local atomic energies $E_i$. $\mathbf{X}_i$ and $A_i$ denote the coordinate and atom type of the *i*-th atom, respectively. (b) In diffusion models, the neural network takes both local environments ($\mathbf{X}_i$ and $A_i$) and the lattice ($\mathbf{L}$) as inputs to score function for coordinates ($\mathbf{s_X}$), atom types ($s_A$), and lattice ($s_L$). Note that the DiffCSP does not predicts score function for atom types, as the chemical formula and atom types are fixed in DiffCSP. In addition, MatterGen uses normalized lattice vectors ($\hat{\mathbf{L}}$) as network inputs, whereas DiffCSP uses the absolute lattice vectors ($\mathbf{L}$).

To evaluate performance in a fixed-composition setting, we train a formula-fixed version of MatterGen on the MP-20 database. In Figure 5j, we test this MatterGen and EquiCSP on $CuAlO_2$, where the experimental structure is included in the training set. We generate 128 structures for each $N_{atom}$, ranging from 4 to 40. The formula-fixed MatterGen and EquiCSP reproduce the ground-state structure up to $N_{atom} = 12$ and $N_{atom} = 16$, respectively, but fails to do so beyond these sizes. Finally, we test DiffCSP on diverse solid-state electrolyte structures (Figure 5k). For each material, we generate 1024 structures at the same formula units as the primitive cell. For $LiNbOCl_4$ ($N_{atom} = 28$), DiffCSP successfully predicts the experimentally known structure despite its absence from the MPTS-52 training set. In contrast, for $Li_2Mg_2S_3O_{12}$ ($N_{atom} = 76$), $LiAlSiO_4$ ($N_{atom} = 84$), and $Li_7La_3Zr_2O_{12}$ ($N_{atom} = 96$), DiffCSP fails to reproduce ground states and generates only high-energy structures.

These results indicate that diffusion models extrapolate poorly for untrained ranges of $N_{atom}$. One might argue that this behavior reflects a simple out-of-distribution (OOD) phenomenon; however, we contend that it represents a more fundamental issue. From an architectural perspective, diffusion models do not predict structures based on global representations. Instead, they operate as per-atom models that predict atomic updates from local environments. Consequently, from this standpoint, an increase in the $N_{\text{atom}}$ does not, by itself, place the system in an extrapolation regime (see Scheme 3a). In fact, this architecture is same as MLIP (Scheme 3b), where the number of atoms in the cell does not affect how the model operates. The model back bone structure is also similar to those used in MLIPs (GemNet[49] and EGNN[50]). Moreover, MLIPs and diffusion models are trained on similar datasets (e.g., Materials Project[2] and Alexandria[38] database). Despite similar training data and architectures, MLIPs often extrapolate successfully to unseen structures and compositions,[27,29] whereas diffusion models often fail dramatically when extrapolating to larger unit cells and to unseen compositions. This qualitative discrepancy suggests that the poor extrapolation of diffusion models cannot be attributed solely to atom-count OOD effects in the conventional sense.

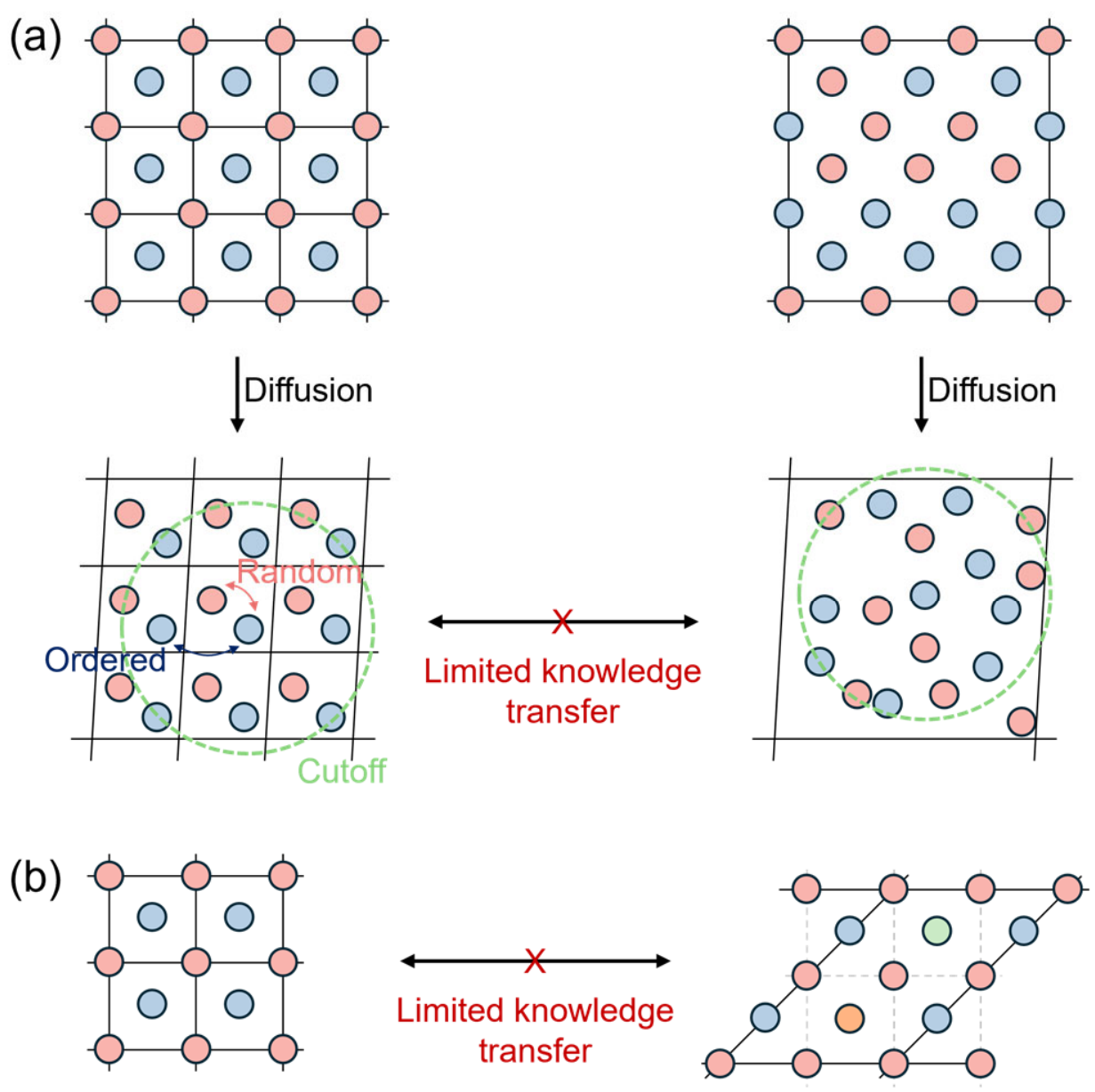

**Scheme 4.** Schematic illustration why diffusion models poorly extrapolate with respect to the number of atoms in the cell.

We identify two main reasons for this difference (Scheme 4). As shown in Scheme 4a, diffusion models operate under periodic boundary conditions. If two structures are similar but differ in lattice parameters, so that one lattice is larger than the other, the corresponding noised structures can have distinct characteristics. With small lattices, diffusion yields noised structures, but they remain relatively ordered due to the periodic boundary condition. In contrast, with larger lattices, the noised structures appear closer to a random distribution of atoms. Because diffusion models learn the relationship between the distributions of such noised structures and the original structures, the distribution of noised samples strongly depends on the system size. Consequently, the model implicitly learns size-specific distributions, making it difficult to connect knowledge across different cell sizes. This mismatch undermines the transferability of diffusion models. Another limitation arises because DiffCSP, EquiCSP, and MatterGen incorporate lattice information directly or indirectly into the score function model (Scheme 4b). As a result, even when the atomic arrangements are similar, their mathematical representations differ, which hinders knowledge transfer across systems.

To theoretically substantiate this interpretation, we demonstrate that the Kullback-Leibler divergence between the distributions of ordered crystal structures $p(\mathbf{x}_0)$ and the noise distribution $p(\mathbf{x}_T)$ in diffusion models is expressed as follows (see Supplementary Information for detailed mathematical demonstration):

$$D\big(p(\mathbf{x}_0)\|p(\mathbf{x}_T)\big) \approx \frac{3N_{\text{atom}}}{3\sigma^2}.$$

Here, we use simple assumption that the distribution of the noised structure follows the mixture of Gaussian distributions, where $\sigma$ is their variance. This indicates that the difference in the data distribution between the noise and crystal structures is a function of $N_{\text{atom}}$ for this simply approximated case, where the real case scenarios would show much more complicated dependence on $N_{\text{atom}}$. Consequently, the data distribution itself changes with the number of atoms, leading to poor extrapolation for larger systems. This differs from conventional OOD scenarios, in which errors arise from insufficient sampling within an otherwise fixed data distribution.

As these phenomena are attributed to the periodic boundary conditions of crystal structures, we term this limitation the *curse of periodicity*. This periodic-boundary-condition–limited knowledge transfer not only restricts extrapolation with respect to the number of atoms in the unit cell, but may also lead to poorer extrapolation in chemical space compared to MLIPs. Interestingly, the extrapolation capability of MatterGen models is better in the Alex-MP-20–trained model than in the MP-20–trained model (Figures 5a,b,d,e), even though both training sets consist only of cells with fewer than 20 atoms. This suggests that enlarging the training set to sample more diverse local geometries can improve the model's coverage of large-cell disordered structures that emerge during the denoising process. To further investigate the role of lattice parameters in the diffusion models, we modify MatterGen by removing the lattice-related term from the model and retraining it. However, consistent with the observations reported in the original paper,[22] the modified model exhibits larger training loss values (see Table S2). In addition, we train two MatterGen models with MP-20 database, where one model is trained on the primitive cell configurations, and the other one is trained on the conventional cell configurations (see Figure S2). We find that the model trained on the conventional cells shows better extrapolation capability, which further supports that the periodicity affects the training quality of diffusion models.

## DISCUSSION

So far, we have tested diffusion models on diverse dataset, including ternary metal oxide, ternary nitride, and GNoME database. We find that the diffusion models perform stably on the oxide and nitride systems. As the generation time for MatterGen for 1024 structures is about 4 hours with an NVIDIA A6000 GPU, and for DiffCSP is about 1~4 hours for generating 4048 structures. The structural optimizations and evaluations with MLIPs and DFT calculations are about total 1~3 hours. In total, it is few tens of times faster than the genetic algorithm with MLIPs. Therefore, we think that the diffusion models can be an effective alternative of the conventional CSP methods in these well-trained chemical spaces.

On the other hand, the performances of these models are relatively poor when tested on the GNoME database, which contains uncommon elements and chemical formulas. The poor performance of diffusion models in under-explored chemical space indicates that diffusion models need to be further developed to embrace a wider chemical space. We believe that an iterative refinement strategy with direct search methods, such as genetic algorithms, is required to provide training set to diffusion models on OOD domains. However, as noted above, while genetic algorithms are generally reliable, it can be inaccurate for some cases due to the limited training of bespoke MLIPs (e.g., those trained with amorphous structures[17]) on diverse local motifs. We suggest that this limitation can be addressed by employing universal MLIPs, which offer broader transferability and wider training coverage than diffusion models. Universal MLIPs can be combined with genetic algorithms as structure generators, and the generated structures can then be re-calculated by DFT calculations to provide additional training data for refining bespoke MLIP models. Alternatively, universal MLIPs can be directly integrated with CSP methods, using them as energy evaluators.[51] Since OOD issues in MLIPs have already been extensively studied, such universal MLIPs can be improved on-the-fly for OOD data through fine-tuning.[52] By contrast, diffusion models pose a more fundamental challenge. As probabilistic generative models, it is inherently difficult to quantify their uncertainty. Therefore, it will be an important research direction to adapt concepts from other machine learning domains to define and assess uncertainty and OOD behavior in diffusion models (for instance: refs. [53–55]).

We introduce the phenomenon we call the *curse of periodicity*, which indicates that diffusion models fail to extrapolate beyond the number of atoms included in the training range. Therefore, for applications such as solid-state electrolytes (e.g., $Li_7La_3Zr_2O_{12}$; 96 atoms in the primitive cell and 192 atoms in the conventional cell), it is challenging to directly apply diffusion models. We consider this phenomenon to be universal, as it is unavoidable that mathematical differences arise when representing crystals with different numbers of atoms under periodic boundary conditions. However, the severity of this issue may vary depending on the model architecture.

For example, in addition to direct diffusion in 3D space, AiDT have been developed that encode crystal and molecular information into a latent space, where diffusion is then performed.[56] In such models, crystallinity may be better captured in the latent representation, potentially reducing the dependence on explicit lattice information. However, these models often neglect rotational symmetry and instead rely on data augmentation, leaving open questions about whether they can match the accuracy of direct-space diffusion models in producing true ground-state structures. Rigorous testing remains necessary, as conducted in this study. Furthermore, it remains unclear whether practical training is possible when including much larger crystal databases, even with data augmentation, underscoring the need for further work. Other directions have also emerged. For instance, a recent study develops a VAE framework that encodes Wyckoff sites and symmetry information into the latent space and reconstructs crystal structures from these representations (named WyCryst).[57] In another line, DiffCSP++ extends DiffCSP by explicitly imposing space-group symmetries during diffusion.[58] SymmCD similarly incorporates symmetry information in the diffusion and denoising process.[59] These symmetry-encoding approaches significantly reduces the search space and stabilizes training, thereby suggesting that they might enable the handling of larger structures. Additionally, language-model-based structure generation approaches have been proposed.[60,61] However, these symmetry-aware and language-model-based methods also require further study to assess their ability to generate energetically stable crystal structures. Overall, while a variety of alternative architectures have been proposed, systematic tests are still required to evaluate their scalability and reliability in materials design.

In addition, based on our theoretical analysis, we suggest directions for improving diffusion models that operate directly on real-space atomic coordinates. As discussed above, a key limitation of existing diffusion models is that they do not account for periodicity-induced differences in the statistics of noised structures for different number of atoms. In current approaches, noised structures arising from unit cells of different sizes are treated as having the same probability distribution, even though their statistical characteristics differ substantially due to periodic boundary conditions. One possible revision is to introduce a periodicity-aware diffusion scheme, in which the noise magnitude is explicitly scaled with lattice size. An alternative strategy is to adapt the number of diffusion steps to the unit-cell size. In addition, providing more explicit lattice information to the diffusion model for the score function may further improve scalability. In MatterGen, the score function takes normalized lattice vector (lattice shape information) as an input, while absolute lattice lengths are not explicitly encoded. Extending the model to include full lattice-vector information may help the network better distinguish between cells of different sizes. In DiffCSP, lattice information is explicitly included; however, during diffusion, all noised lattice configurations are driven toward an identical isotropic Gaussian distribution, effectively shrinking cells in a size-independent manner (the average lattice parameter of the noise structures is 1 Å for all $N_{atom}$). From a per-atom perspective, this leads to unintended variations in atomic density. Because diffusion models learn score functions on a per-atom basis (Scheme 4b), we argue that preserving atomic density during diffusion is more natural, as it in MatterGen. We therefore suggest that incorporating density-preserving lattice diffusion in DiffCSP into future models could substantially improve their scalability to larger unit cells. However, we note that such modifications may also limit transferability across different lattice sizes, as discussed above with Scheme 3b, highlighting the importance of maintaining a proper balance.

Finally, we want to mention that there are only a few direct search methods capable of reliably handling complex structures with large number of atoms, including quaternary and higher compositions with more than 100 atoms. Accordingly, datasets for training diffusion models on such crystal structures remain scarce. Recently, Han et al. introduced a symmetry-guided search strategy that improves performance for complex systems.[62] This method is validated for crystal structures such as $Mg_3Al_2Si_3O_{12}$ ($N_{atom}$ = 80 for primitive cell and $N_{atom}$ = 160 for conventional cell). However, although $N_{atom}$ is large, this material has relatively simple symmetry. Indeed, in ref. [63], it was found that $LiTa_2PO_8$ ($N_{atom}$ = 48 for primitive cell and $N_{atom}$ = 96 for conventional cell) is about 5,000 times more difficult to identify than $Na_3Ga_3Te_2O_{12}$ (same structure as $Mg_3Al_2Si_3O_{12}$), despite having fewer atoms but a more complex structure. This highlights the need for further validation of the method on other multinary systems with large number of atoms. In addition, a recent study developed an integer–programming–based search method that guarantees optimality and enables rapid identification of large-atom structures.[64] However, this method oversimplifies the potential energy surface using a binary quadratic form and does not explore optimal lattice vectors. Thus, its generalization to broader material families remains unproven. Lastly, a structure prediction method has been proposed for targeting oxide solid electrolytes by giving constraints on corner-sharing bond topologies.[63] While this approach has been well validated across several systems, such as $LiTa_2PO_8$ and $Li_7La_3Zr_2O_{12}$ ($N_{atom}$ = 96 for primitive cell and $N_{atom}$ = 192 for conventional cell), it is restricted to frameworks with corner-sharing motifs. Overall, continued efforts to develop direct search methods are essential for constructing datasets suitable for training diffusion models on these challenging systems.

## CONCLUSION

In summary, we systematically evaluate the capability of diffusion models, MatterGen and DiffCSP, for crystal structure prediction across different chemical spaces and system sizes. Our results demonstrate that diffusion models perform reliably in well-sampled chemical spaces such as ternary oxides and nitrides, but their performance deteriorates significantly in uncommon compositions, as exemplified by the GNoME database containing rare-earth elements and unconventional stoichiometries. Moreover, we reveal that diffusion models suffer from poor size extrapolation, with accuracy dropping significantly once the number of atoms exceeds the trained range. We attribute this limitation to the periodic boundary conditions inherent to crystalline systems, which we term the *curse of periodicity*. These findings highlight both the strengths and the limitations of diffusion models for materials discovery. While diffusion models offer a promising pathway for efficient crystal structure generation, their lack of robustness in underrepresented chemical spaces and large-cell systems underscores the need for methodological advances.

## METHODS

### Diffusion models

We use the pretrained models for the tests on the ternary metal oxide, ternary nitride, and GNoME databases. The diffusion guidance factor is set to 2.0, and the batch size is fixed at 1024. For the MatterGen model trained on the MP-20 database, we follow the official MatterGen package and adopt the same hyperparameters as those used in the Alex-MP-20 pretrained model. To evaluate the size extrapolation capability, we generate 100 structures for each fixed $N_{atom}$ in systems with up to six components. To analyze the elemental distributions in the structures generated by MP-20-trained MatterGen, Alex-MP-20-trained MatterGen, and MP-20-trained EquiCSP, we generate 128 structures for each fixed $N_{atom}$ with a batch size of 32. The chemical-formula-fixed MatterGen model for Figure 5j is trained in CSP mode with atom and edge embedding dimensions of 128 (the pretrained model uses 512). The target composition of the generated structures is $CuAlO_2$, and for each fixed $N_{atom}$ from 4 to 40, we generate 128 structures with a batch size of 32. The energies of these structures are also calculated using SevenNet-MF-ompa. For DiffCSP, we use the pretrained model trained on the MPTS-52 database, with the batch size of 1024 per formula unit.

### SPINNER

The training set generation procedure is done automatic MD tools following the procedure in ref. 18. In structure search, structures are generated using random seeding (70%), permutation (20%), and lattice mutation (10%). Each generation contains up to 300 structures, and the search continues for 200 generations. In every generation, structures lying within 100 meV/atom of the lowest-energy structure from the previous generation are chosen for mutation. After completion, the final candidate structures are selected from those within 50 meV/atom of the lowest-energy structure, followed by DFT calculations.

### Optimization and energy evaluations

For the ternary metal oxide and ternary nitride databases, we first optimize (converged to 0.05 eV/Å) the generated structures using SevenNet-MF-ompa with the *mpa* modal (consistent with the Materials Project and Alexandria databases). The structures within a 30 meV/atom hull-energy window are then recalculated with DFT. For the GNoME tests, we first perform single-shot calculations and select structures within a 60 meV/atom hull-energy window, which are subsequently fully relaxed. For structures with the same chemical formula as the GNoME structures, we calculate the energy difference between the GNoME and predicted structures and obtain the hull energy of the predicted structures by adding this difference to the reported hull energy of the corresponding GNoME structure. For structures with different chemical formulas, we calculate the hull energies by referencing the Materials Project database, as in the oxide and nitride tests. DFT calculations are performed only when the energy obtained by SevenNet-MF-ompa is at least 50 meV/atom lower than the corresponding GNoME structures. Note that the standard pseudopotential for Yb in the Materials Project has been changed, which causes the pymatgen phase diagram and correction modules to fail in evaluating the hull energies of Yb-containing compounds. Therefore, we exclude these species when the chemical formula of the predicted structures differs from that of the GNoME structures.

All DFT calculations are performed similarly to the Materials Project protocol. We use the VASP code[65] with projector augmented-wave (PAW) pseudopotentials.[66] The PBE exchange–correlation functional[44] is employed. Cutoff energies and *k*-point grids are determined by convergence tests with thresholds of 2 meV/atom for energy, 0.1 eV/Å for force, and 10 kbar for stress. The minimum cutoff energy is set to 500 eV. Cutoff energy tests, k-point sampling, and structural optimizations are performed automatically using AMP[2].[41]

### Visualization of atomic structures

All atomic configurations in this paper are drawn with the VESTA code.[67]

## RESOURCE AVAILABILITY

### Lead contact

Requests for further information and resources should be directed to and will be fulfilled by the lead contact, Sungwoo Kang (sung.w.kang@kist.re.kr).

### Data availability

All stable structures discovered for each chemical formula or system, along with their corresponding energies, are available at: https://github.com/kang1717/diffusion_model_test_data/.

## ACKNOWLEDGEMENTS

This work was supported by the Nano and Material Technology Development Programs through the National Research Foundation of Korea (NRF) funded by the Ministry of Science and ICT (Grant No. RS-2024-00407995 and No. RS-2024-00450102).

## AUTHOR CONTRIBUTIONS

S.Ki. and G.J. contributed equally. S.Ki. performed comparative tests between diffusion models and CSP methods. G.J. conducted scalability tests with respect to the number of atoms in the unit cell. G.J. and S.Ka. developed the theory of the curse of periodicity. G.J. provided the mathematical demonstration of the curse of periodicity. S.Hw., J.L., J.J., and S.Ha. carried out the comparison between SPINNER and GNoME. S.Ka. supervised the overall project. All authors contributed to writing the manuscript.

## DECLARATION OF INTERESTS

The authors declare no competing interests.

## References

(1) Bergerhoff, G.; Hundt, R.; Sievers, R.; Brown, I. D. The Inorganic Crystal Structure Data Base. *J. Chem. Inf. Comput. Sci.* **1983**, *23* (2), 66–69. https://doi.org/10.1021/ci00038a003.

(2) Jain, A.; Ong, S. P.; Hautier, G.; Chen, W.; Richards, W. D.; Dacek, S.; Cholia, S.; Gunter, D.; Skinner, D.; Ceder, G.; Persson, K. A. Commentary: The Materials Project: A Materials Genome Approach to Accelerating Materials Innovation. *APL Materials* **2013**, *1* (1), 011002. https://doi.org/10.1063/1.4812323.

(3) Oganov, A. R.; Pickard, C. J.; Zhu, Q.; Needs, R. J. Structure Prediction Drives Materials Discovery. *Nat Rev Mater* **2019**, *4* (5), 331–348. https://doi.org/10.1038/s41578-019-0101-8.

(4) Maddox, J. Crystals from First Principles. *Nature* **1988**, *335*, 201.

(5) Pickard, C. J.; Needs, R. J. High-Pressure Phases of Silane. *Phys. Rev. Lett.* **2006**, *97* (4), 045504. https://doi.org/10.1103/PhysRevLett.97.045504.

(6) Glass, C. W.; Oganov, A. R.; Hansen, N. USPEX—Evolutionary Crystal Structure Prediction. *Computer Physics Communications* **2006**, *175* (11–12), 713–720. https://doi.org/10.1016/j.cpc.2006.07.020.

(7) Wang, Y.; Lv, J.; Zhu, L.; Ma, Y. Crystal Structure Prediction via Particle-Swarm Optimization. *Phys. Rev. B* **2010**, *82* (9), 094116. https://doi.org/10.1103/PhysRevB.82.094116.

(8) Fischer, C. C.; Tibbetts, K. J.; Morgan, D.; Ceder, G. Predicting Crystal Structure by Merging Data Mining with Quantum Mechanics. *Nature Mater* **2006**, *5* (8), 641–646. https://doi.org/10.1038/nmat1691.

(9) Davies, D. W.; Butler, K. T.; Skelton, J. M.; Xie, C.; Oganov, A. R.; Walsh, A. Computer-Aided Design of Metal Chalcohalide Semiconductors: From Chemical Composition to Crystal Structure. *Chem. Sci.* **2018**, *9* (4), 1022–1030. https://doi.org/10.1039/C7SC03961A.

(10) Merchant, A.; Batzner, S.; Schoenholz, S. S.; Aykol, M.; Cheon, G.; Cubuk, E. D. Scaling Deep Learning for Materials Discovery. *Nature* **2023**, *624* (7990), 80–85. https://doi.org/10.1038/s41586-023-06735-9.

(11) Behler, J. First Principles Neural Network Potentials for Reactive Simulations of Large Molecular and Condensed Systems. *Angew Chem Int Ed* **2017**, *56* (42), 12828–12840. https://doi.org/10.1002/anie.201703114.

(12) Bartók, A. P.; Payne, M. C.; Kondor, R.; Csányi, G. Gaussian Approximation Potentials: The Accuracy of

Quantum Mechanics, without the Electrons. *Phys. Rev. Lett.* **2010**, *104* (13), 136403. https://doi.org/10.1103/PhysRevLett.104.136403.

(13) Unke, O. T.; Chmiela, S.; Sauceda, H. E.; Gastegger, M.; Poltavsky, I.; Schütt, K. T.; Tkatchenko, A.; Müller, K.-R. Machine Learning Force Fields. *Chem. Rev.* **2021**, *121* (16), 10142–10186. https://doi.org/10.1021/acs.chemrev.0c01111.

(14) Deringer, V. L.; Pickard, C. J.; Csányi, G. Data-Driven Learning of Total and Local Energies in Elemental Boron. *Phys. Rev. Lett.* **2018**, *120* (15), 156001. https://doi.org/10.1103/PhysRevLett.120.156001.

(15) Podryabinkin, E. V.; Tikhonov, E. V.; Shapeev, A. V.; Oganov, A. R. Accelerating Crystal Structure Prediction by Machine-Learning Interatomic Potentials with Active Learning. *Phys. Rev. B* **2019**, *99* (6), 064114. https://doi.org/10.1103/PhysRevB.99.064114.

(16) Tong, Q.; Xue, L.; Lv, J.; Wang, Y.; Ma, Y. Accelerating CALYPSO Structure Prediction by Data-Driven Learning of a Potential Energy Surface. *Faraday Discuss.* **2018**, *211*, 31–43. https://doi.org/10.1039/C8FD00055G.

(17) Kang, S.; Jeong, W.; Hong, C.; Hwang, S.; Yoon, Y.; Han, S. Accelerated Identification of Equilibrium Structures of Multicomponent Inorganic Crystals Using Machine Learning Potentials. *npj Comput. Mater.* **2022**, *8* (1), 108. https://doi.org/10.1038/s41524-022-00792-w.

(18) Hwang, S.; Jung, J.; Hong, C.; Jeong, W.; Kang, S.; Han, S. Stability and Equilibrium Structures of Unknown Ternary Metal Oxides Explored by Machine-Learned Potentials. *J. Am. Chem. Soc.* **2023**, *145* (35), 19378–19386. https://doi.org/10.1021/jacs.3c06210.

(19) Noh, J.; Kim, J.; Stein, H. S.; Sanchez-Lengeling, B.; Gregoire, J. M.; Aspuru-Guzik, A.; Jung, Y. Inverse Design of Solid-State Materials via a Continuous Representation. *Matter* **2019**, *1* (5), 1370–1384. https://doi.org/10.1016/j.matt.2019.08.017.

(20) Xie, T.; Fu, X.; Ganea, O.-E.; Barzilay, R.; Jaakkola, T. Crystal Diffusion Variational Autoencoder for Periodic Material Generation. arXiv March 14, 2022. https://doi.org/10.48550/arXiv.2110.06197.

(21) Jiao, R.; Huang, W.; Lin, P.; Han, J.; Chen, P.; Lu, Y.; Liu, Y. Crystal Structure Prediction by Joint Equivariant Diffusion. arXiv March 7, 2024. https://doi.org/10.48550/arXiv.2309.04475.

(22) Zeni, C.; Pinsler, R.; Zügner, D.; Fowler, A.; Horton, M.; Fu, X.; Wang, Z.; Shysheya, A.; Crabbé, J.; Ueda, S.; Sordillo, R.; Sun, L.; Smith, J.; Nguyen, B.; Schulz, H.; Lewis, S.; Huang, C.-W.; Lu, Z.; Zhou, Y.; Yang, H.; Hao, H.; Li, J.; Yang, C.; Li, W.; Tomioka, R.; Xie, T. A Generative Model for Inorganic Materials Design. *Nature* **2025**, *639* (8055), 624–632. https://doi.org/10.1038/s41586-025-08628-5.

(23) Takamoto, S.; Shinagawa, C.; Motoki, D.; Nakago, K.; Li, W.; Kurata, I.; Watanabe, T.; Yayama, Y.; Iriguchi, H.; Asano, Y.; Onodera, T.; Ishii, T.; Kudo, T.; Ono, H.; Sawada, R.; Ishitani, R.; Ong, M.; Yamaguchi, T.; Kataoka, T.; Hayashi, A.; Charoenphakdee, N.; Ibuka, T. Towards Universal Neural Network Potential for Material Discovery Applicable to Arbitrary Combination of 45 Elements. *Nat Commun* **2022**, *13* (1), 2991. https://doi.org/10.1038/s41467-022-30687-9.

(24) Chen, C.; Ong, S. P. A Universal Graph Deep Learning Interatomic Potential for the Periodic Table. *Nat Comput Sci* **2022**, *2* (11), 718–728. https://doi.org/10.1038/s43588-022-00349-3.

(25) Deng, B.; Zhong, P.; Jun, K.; Riebesell, J.; Han, K.; Bartel, C. J.; Ceder, G. CHGNet as a Pretrained Universal Neural Network Potential for Charge-Informed Atomistic Modelling. *Nat Mach Intell* **2023**, *5* (9), 1031–1041. https://doi.org/10.1038/s42256-023-00716-3.

(26) Batatia, I.; Benner, P.; Chiang, Y.; Elena, A. M.; Kovács, D. P.; Riebesell, J.; Advincula, X. R.; Asta, M.; Avaylon, M.; Baldwin, W. J.; Berger, F.; Bernstein, N.; Bhowmik, A.; Blau, S. M.; Cărare, V.; Darby, J. P.; De, S.; Della Pia, F.; Deringer, V. L.; Elijošius, R.; El-Machachi, Z.; Falcioni, F.; Fako, E.; Ferrari, A. C.; Genreith-Schriever, A.; George, J.; Goodall, R. E. A.; Grey, C. P.; Grigorev, P.; Han, S.; Handley, W.; Heenen, H. H.; Hermansson, K.; Holm, C.; Jaafar, J.; Hofmann, S.; Jakob, K. S.; Jung, H.; Kapil, V.; Kaplan, A. D.; Karimitari, N.; Kermode, J. R.; Kroupa, N.; Kullgren, J.; Kuner, M. C.; Kuryla, D.; Liepuoniute, G.; Margraf, J. T.; Magdău, I.-B.; Michaelides, A.; Moore, J. H.; Naik, A. A.; Niblett, S. P.; Norwood, S. W.; O'Neill, N.; Ortner, C.; Persson, K. A.; Reuter, K.; Rosen, A. S.; Schaaf, L. L.; Schran, C.; Shi, B. X.; Sivonxay, E.; Stenczel, T. K.; Svahn, V.; Sutton, C.; Swinburne, T. D.; Tilly, J.; van der Oord, C.; Varga-Umbrich, E.; Vegge, T.; Vondrák, M.; Wang, Y.; Witt, W. C.; Zills, F.; Csányi, G. A Foundation Model for Atomistic Materials Chemistry. arXiv March 1, 2024. http://arxiv.org/abs/2401.00096 (accessed 2024-07-16).

(27) Kang, S. How Graph Neural Network Interatomic Potentials Extrapolate: Role of the Message-Passing Algorithm. *J. Chem. Phys.* **2024**, *161* (24), 244102. https://doi.org/10.1063/5.0234287.

(28) Zhang, Y.-W.; Sorkin, V.; Aitken, Z. H.; Politano, A.; Behler, J.; P Thompson, A.; Ko, T. W.; Ong, S. P.; Chalykh, O.; Korogod, D.; Podryabinkin, E.; Shapeev, A.; Li, J.; Mishin, Y.; Pei, Z.; Liu, X.; Kim, J.; Park, Y.; Hwang, S.; Han, S.; Sheriff, K.; Cao, Y.; Freitas, R. Roadmap for the Development of Machine Learning-Based Interatomic Potentials. *Modelling Simul. Mater. Sci. Eng.* **2025**, *33* (2), 023301. https://doi.org/10.1088/1361-651X/ad9d63.

(29) Riebesell, J.; Goodall, R. E. A.; Benner, P.; Chiang, Y.; Deng, B.; Ceder, G.; Asta, M.; Lee, A. A.; Jain, A.; Persson, K. A. A Framework to Evaluate Machine Learning Crystal Stability Predictions. *Nat Mach Intell* **2025**, *7* (6), 836–847. https://doi.org/10.1038/s42256-025-01055-1.

(30) Yang, H.; Hu, C.; Zhou, Y.; Liu, X.; Shi, Y.; Li, J.; Li, G.; Chen, Z.; Chen, S.; Zeni, C.; Horton, M.; Pinsler, R.; Fowler, A.; Z¨ugner, D.; Xie, T.; Smith, J.; Sun, L.; Wang, Q.; Kong, L.; Liu, C.; Hao, H.; Lu, Z. MatterSim: A Deep Learning Atomistic Model Across Elements, Temperatures and Pressures. *arXiv:2405.04967v2* **2024**.

(31) Yeo, B. C.; Kang, S.; Lee, J.-H. Diffusion–Model–Driven Discovery of Ferroelectrics for Photocurrent Applications. Chemistry September 26, 2025. https://doi.org/10.26434/chemrxiv-2025-qrm65.

(32) Park, H.; Li, Z.; Walsh, A. Has Generative Artificial Intelligence Solved Inverse Materials Design? *Matter* **2024**, *7* (7), 2355–2367. https://doi.org/10.1016/j.matt.2024.05.017.

(33) Sun, W.; Dacek, S. T.; Ong, S. P.; Hautier, G.; Jain, A.; Richards, W. D.; Gamst, A. C.; Persson, K. A.; Ceder, G.

The Thermodynamic Scale of Inorganic Crystalline Metastability. *Sci. Adv.* **2016**, *2* (11), e1600225. https://doi.org/10.1126/sciadv.1600225.

(34) Negishi, M.; Park, H.; Mastej, K. O.; Walsh, A. Continuous Uniqueness and Novelty Metrics for Generative Modeling of Inorganic Crystals. arXiv October 23, 2025. https://doi.org/10.48550/arXiv.2510.12405.

(35) Lin, P.; Chen, P.; Jiao, R.; Mo, Q.; Cen, J.; Huang, W.; Liu, Y.; Huang, D.; Lu, Y. Equivariant Diffusion for Crystal Structure Prediction. arXiv December 8, 2025. https://doi.org/10.48550/arXiv.2512.07289.

(36) Yang, S.; Cho, K.; Merchant, A.; Abbeel, P.; Schuurmans, D.; Mordatch, I.; Cubuk, E. D. Scalable Diffusion for Materials Generation. arXiv June 3, 2024. https://doi.org/10.48550/arXiv.2311.09235.

(37) Sun, W.; Bartel, C. J.; Arca, E.; Bauers, S. R.; Matthews, B.; Orvañanos, B.; Chen, B.-R.; Toney, M. F.; Schelhas, L. T.; Tumas, W.; Tate, J.; Zakutayev, A.; Lany, S.; Holder, A. M.; Ceder, G. A Map of the Inorganic Ternary Metal Nitrides. *Nat. Mater.* **2019**, *18* (7), 732–739. https://doi.org/10.1038/s41563-019-0396-2.

(38) Schmidt, J.; Wang, H.-C.; Cerqueira, T. F. T.; Botti, S.; Marques, M. A. L. A Dataset of 175k Stable and Metastable Materials Calculated with the PBEsol and SCAN Functionals. *Sci Data* **2022**, *9* (1), 64. https://doi.org/10.1038/s41597-022-01177-w.

(39) Baird, S. G.; Sayeed, H. M.; Montoya, J.; Sparks, T. D. Matbench-Genmetrics: A Python Library for Benchmarkingcrystal Structure Generative Models Using Time-Based Splits of MaterialsProject Structures. *JOSS* **2024**, *9* (97), 5618. https://doi.org/10.21105/joss.05618.

(40) Kim, J.; Kim, J.; Kim, J.; Lee, J.; Park, Y.; Kang, Y.; Han, S. Data-Efficient Multifidelity Training for High-Fidelity Machine Learning Interatomic Potentials. *J. Am. Chem. Soc.* **2025**, *147* (1), 1042–1054. https://doi.org/10.1021/jacs.4c14455.

(41) Youn, Y.; Lee, M.; Hong, C.; Kim, D.; Kim, S.; Jung, J.; Yim, K.; Han, S. AMP2: A Fully Automated Program for Ab Initio Calculations of Crystalline Materials. *Computer Physics Communications* **2020**, *256*, 107450. https://doi.org/10.1016/j.cpc.2020.107450.

(42) Hong, C.; Choi, J. M.; Jeong, W.; Kang, S.; Ju, S.; Lee, K.; Jung, J.; Youn, Y.; Han, S. Training Machine-Learning Potentials for Crystal Structure Prediction Using Disordered Structures. *Phys. Rev. B* **2020**, *102* (22), 224104. https://doi.org/10.1103/PhysRevB.102.224104.

(43) Furness, J. W.; Kaplan, A. D.; Ning, J.; Perdew, J. P.; Sun, J. Accurate and Numerically Efficient $r^2$ SCAN Meta-Generalized Gradient Approximation. *J. Phys. Chem. Lett.* **2020**, *11* (19), 8208–8215. https://doi.org/10.1021/acs.jpclett.0c02405.

(44) Perdew, J. P.; Burke, K.; Ernzerhof, M. Generalized Gradient Approximation Made Simple. *Phys. Rev. Lett.* **1996**, *77* (18), 3865–3868. https://doi.org/10.1103/PhysRevLett.77.3865.

(45) Hicks, D.; Toher, C.; Ford, D. C.; Rose, F.; Santo, C. D.; Levy, O.; Mehl, M. J.; Curtarolo, S. AFLOW-XtalFinder: A Reliable Choice to Identify Crystalline Prototypes. *npj Comput Mater* **2021**, *7* (1), 30. https://doi.org/10.1038/s41524-020-00483-4.

(46) Cheetham, A. K.; Seshadri, R. Artificial Intelligence Driving Materials Discovery? Perspective on the Article: Scaling Deep Learning for Materials Discovery. *Chem. Mater.* **2024**, *36* (8), 3490–3495. https://doi.org/10.1021/acs.chemmater.4c00643.

(47) Togo, A.; Chaput, L.; Tadano, T.; Tanaka, I. Implementation Strategies in Phonopy and Phono3py. *J. Phys.: Condens. Matter* **2023**, *35* (35), 353001. https://doi.org/10.1088/1361-648X/acd831.

(48) Togo, A. First-Principles Phonon Calculations with Phonopy and Phono3py. *J. Phys. Soc. Jpn.* **2023**, *92* (1), 012001. https://doi.org/10.7566/JPSJ.92.012001.

(49) Gasteiger, J.; Becker, F.; Günnemann, S. GemNet: Universal Directional Graph Neural Networks for Molecules. arXiv April 5, 2022. http://arxiv.org/abs/2106.08903 (accessed 2024-04-17).

(50) Satorras, V. G.; Hoogeboom, E.; Welling, M. E(n) Equivariant Graph Neural Networks. arXiv February 16, 2022. https://doi.org/10.48550/arXiv.2102.09844.

(51) Shibayama, T.; Imamura, H.; Nishimra, K.; Shinohara, K.; Shinagawa, C.; Takamoto, S.; Li, J. Efficient Crystal Structure Prediction Using Genetic Algorithm and Universal Neural Network Potential. arXiv June 24, 2025. https://doi.org/10.48550/arXiv.2503.21201.

(52) Kim, J.; Lee, J.; Oh, S.; Park, Y.; Hwang, S.; Han, S.; Kang, S.; Kang, Y. An Efficient Forgetting-Aware Fine-Tuning Framework for Pretrained Universal Machine-Learning Interatomic Potentials. arXiv June 18, 2025. https://doi.org/10.48550/arXiv.2506.15223.

(53) Heng, A.; Thiery, A. H.; Soh, H. Out-of-Distribution Detection with a Single Unconditional Diffusion Model.

(54) Lu, S.; Wang, Y.; Sheng, L.; He, L.; Zheng, A.; Liang, J. Out-of-Distribution Detection: A Task-Oriented Survey of Recent Advances. arXiv August 4, 2025. https://doi.org/10.48550/arXiv.2409.11884.

(55) Ding, Y.; Aleksandraus, A.; Ahmadian, A.; Unger, J.; Lindsten, F.; Eilertsen, G. Revisiting Likelihood-Based Out-of-Distribution Detection by Modeling Representations. arXiv July 10, 2025. https://doi.org/10.48550/arXiv.2504.07793.

(56) Joshi, C. K.; Fu, X.; Liao, Y.-L.; Gharakhanyan, V.; Miller, B. K.; Sriram, A.; Ulissi, Z. W. All-Atom Diffusion Transformers: Unified Generative Modelling of Molecules and Materials. arXiv May 22, 2025. https://doi.org/10.48550/arXiv.2503.03965.

(57) Zhu, R.; Nong, W.; Yamazaki, S.; Hippalgaonkar, K. WyCryst: Wyckoff Inorganic Crystal Generator Framework. *Matter* **2024**, *7* (10), 3469–3488. https://doi.org/10.1016/j.matt.2024.05.042.

(58) Jiao, R.; Huang, W.; Liu, Y.; Zhao, D.; Liu, Y. Space Group Constrained Crystal Generation. arXiv April 8, 2024. https://doi.org/10.48550/arXiv.2402.03992.

(59) Levy, D.; Panigrahi, S. S.; Kaba, S.-O.; Zhu, Q.; Lee, K. L. K.; Galkin, M.; Miret, S.; Ravanbakhsh, S. SymmCD: Symmetry-Preserving Crystal Generation with Diffusion Models. arXiv May 23, 2025. https://doi.org/10.48550/arXiv.2502.03638.

(60) Antunes, L. M.; Butler, K. T.; Grau-Crespo, R. Crystal Structure Generation with Autoregressive Large Language Modeling. arXiv February 12, 2024. https://doi.org/10.48550/arXiv.2307.04340.

(61) Mok, D. H.; Back, S. Generative Pretrained Transformer for Heterogeneous Catalysts. *J. Am. Chem. Soc.* **2024**, *146* (49), 33712–33722. https://doi.org/10.1021/jacs.4c11504.

(62) Han, Y.; Ding, C.; Wang, J.; Gao, H.; Shi, J.; Yu, S.; Jia, Q.; Pan, S.; Sun, J. Efficient Crystal Structure Prediction Based on the Symmetry Principle. *Nat Comput Sci* **2025**, *5* (3), 255–267. https://doi.org/10.1038/s43588-025-00775-z.

(63) Hwang, S.; Lee, J.; Han, S.; Kang, Y.; Kang, S. Discovery of Oxide Li-Conducting Electrolytes in Uncharted Chemical Space via Topology-Constrained Crystal Structure Prediction. arXiv October 1, 2025. https://doi.org/10.48550/arXiv.2509.25763.

(64) Gusev, V. V.; Adamson, D.; Deligkas, A.; Antypov, D.; Collins, C. M.; Krysta, P.; Potapov, I.; Darling, G. R.; Dyer, M. S.; Spirakis, P.; Rosseinsky, M. J. Optimality Guarantees for Crystal Structure Prediction. *Nature* **2023**, *619* (7968), 68–72. https://doi.org/10.1038/s41586-023-06071-y.

(65) Kresse, G.; Furthmüller, J. Efficiency of Ab-Initio Total Energy Calculations for Metals and Semiconductors Using a Plane-Wave Basis Set. *Comput. Mater. Sci.* **1996**, *6* (1), 15–50. https://doi.org/10.1016/0927-0256(96)00008-0.

(66) Blöchl, P. E. Projector Augmented-Wave Method. *Phys. Rev. B* **1994**, *50* (24), 17953–17979. https://doi.org/10.1103/PhysRevB.50.17953.

(67) Momma, K.; Izumi, F. *VESTA 3* for Three-Dimensional Visualization of Crystal, Volumetric and Morphology Data. *J. Appl. Crystallogr.* **2011**, *44* (6), 1272–1276. https://doi.org/10.1107/S0021889811038970.